\magnification\magstep1
\tolerance=1600
\parskip = 6pt
\def\dt{{\cdot}}
\def\half{{1\over 2}}
\def\hhalf{\textstyle{1\over 2}}
\def\ref#1{$^{[#1]}$}
\def\pagenumber{\footline={\hss\tenrm\folio\hss}}
\def\nox{{\scriptstyle{\times \atop \times}}}
\nopagenumbers
\rightline{IFP/102/UNC}
\vskip 20pt
\footline={\sevenrm
\hfil$^\ast$ Supported in part by the U.S. Department of
Energy under Grant No. DE-FG 05-85ER40219/Task A\hfil}
\vskip 3pt
\centerline{\bf GAUGE SYMMETRY IN BACKGROUND 
CHARGE CONFORMAL FIELD THEORY$^\ast$}
\vskip 50 pt
\centerline {\bf L. Dolan}
\vskip 12pt
\centerline{\it
Department of Physics and Astronomy,University of North Carolina}
\centerline{\it Chapel Hill, North Carolina 27599-3255, USA}
\vskip 50pt
\centerline {\bf ABSTRACT}
{\rightskip=18 true mm \leftskip=18 true mm  \noindent
We present a mechanism to construct four-dimensional
charged massless Ramond states using the discrete states
of a fivebrane Liouville internal conformal field theory. 
This conformal field theory has background charge, and admits
an inner product which allows positive norm states. 
A connection among supergravity soliton solutions, Liouville
conformal field theory, non-critical string theory and their 
gauge symmetry properties is given. A generalized construction of
the $SU(2)$ super Kac-Moody algebra mixing with the $N=1$ super Virasoro
algebra is analyzed. How these Ramond states evade the DKV no-go theorem is
explained. 
}
\vfill
\centerline{October 1996}
\centerline{Submitted to Nuclear Physics B}
\vskip12pt
\vfill\eject
\pagenumber
\centerline{\bf1. Introduction}
\vskip4pt
 
The descriptions of string theory given by
1) solutions of low-energy classical target space supergravity theories,
including non-perturbative BPS solitons,
2) worldsheet sigma model actions, and 3) conformal field theories
of vertex operators have mutually reinforced our understanding
of the spectrum and interactions of the system.
In particular, duality relations among strong and weak coupling limits of 
various string models, and the presence of BPS non-perturbative 
states,
some of which are massless, 
have signalled the importance of non-perturbative effects in our current
formulation of string theory. 
 
In this paper we analyze aspects of gauge symmetry in string theory 
using conformal field theories which already incorporate non-perturbative
information, i.e. conformal field theories involving a coordinate field
with background charge, such as the Liouville superfield. 
As a definite example,
we consider the magnetically charged fivebrane soliton solution of 
the Type II string and its corresponding conformal field theory with
background charge\ref{1-5}. 
We demonstrate how the role of gauge symmetry differs
in these theories
from the action of gauge symmetry in critical string theory.

In particular, we show how the discrete states of the Liouville theory
can be used to construct four-dimensional massless Ramond states
which carry non-trivial charge under the non-abelian 
gauge symmetry of an $SU(2)$ super Kac-Moody algebra present on
this Ramond side.
We explain how this construction evades the arguments of the 
DKV no-go theorem. 

Conventional superstring spectra given by conformal field theories
can be changed  
by allowing non-vanishing vacuum expectation values (vevs)
for any field in the low-energy supergravity,
rather than only for the metric tensor.
The vevs are given by classical solutions, either elementary
or solitonic, of the
supergravity field equations.
 
Soliton solutions are non-perturbative, and exist in both gauge theories
and string theory.
In particular, in Type II string models,
they lead to states which enhance the original spectrum.
This is interesting in that Type II was viewed
as an economical model, but one which appeared to fall just short
of phenomenological viability.  Now with the inclusion of solitons,
these theories are seen to have as many
states as the heterotic string, by the mere choice of
an appropriate internal space to compactify to four dimensions\ref{6,7}.
In this paper, we investigate an origin of spectrum enhancement
present already in a conformal field theory (CFT) description.

The bosonic supergravity fields, apart from the metric,
are described by d-form potentials, which fall into two categories:
they are either
NS-NS or RR, the latter occurring only in Type II superstrings.
These two types are distinguished by the d-forms representing massless 
states having their origin in the
Neveu-Schwarz/Neveu-Schwarz (NS)
or the Ramond/Ramond (R) sector of the string theory.
 
Since RR-branes couple to RR field potentials, the RR-branes carry charge
of the RR bosons.
They have been dubbed Dirichlet p-branes, as they can be
shown to correspond to states in the conformal field theory 
of type I string theory with Dirichlet boundary conditions\ref{8,9},
where the open string ends live on D-branes.

The fivebrane soliton we study is an NS-NS brane, and it
describes a structure which carries charge under the NS-NS gauge bosons
of the Type II superstring. In a limit corresponding to the 
semi-wormhole throat, the corresponding sigma model
is known to be a superconformal field theory. Here we analyze  
its primary field content 
and the nature of its gauge symmetry currents.  
In sect.2, we briefly review the global $N=4$ 
conformal field theory description of 
the fivebrane soliton and discuss the primary fields of the
$N=1$ subalgebra, which is local.
In sect.3, the degrees of freedom of a $d=4$ string model, 
whose internal degrees of freedom can form global extended worldsheet
symmetries, are given. 
The ghost system has conformal charge $c=-15$ on each side. 
In sect.4, the highest weight states and corresponding 
vertex operators of the $N=4$ superconformal algebra
for both massive and massless (short) representations are computed.  
In sect.5, we give the fivebrane conformal field theory representation 
of the superconformal current algebra, which
describes gauge symmetry in string theory.  
The non-zero charge of massless fermion representations is calculated.
Eq.'s (5.9)-(5.17) contain the precise statement explaining how these 
states can exist. 
Conclusions and comments are found in sect.6. The appendix
contains the operator product algebra of the $N=4$ superconformal theory.

\vskip15pt
\centerline{\bf 2. Conformal field theory with background charge}
\vskip5pt

The solitonic fivebrane solution\ref{1-3}
in sigma model coordinates corresponding to the semi-wormhole throat
has a metric given by  
$$ds^2 = \eta_{MN} dx^M dx^N + ({k_6\over {y^2}} e^{C_0} )\delta_{mn}
dy^m dy^n\eqno (2.1)$$
where $0\le M,N\le 5$, and $6\le m,n\le 9$. 
The first term corresponds to a $c=9$ CFT, and
the second term represents a $c=6$ CFT, where $k_6$ and 
$C_0$ are certain constants which
contain a parameter $k$, the axion charge.
We note that the (M,N) space-time part
of the metric is flat, so that the $c=9$ part is a free 
conformal field theory\ref{1-3}. 

The $c=6$ part is a nontrivial conformal
field theory, and we call it the $c=6$ 
``fivebrane'' conformal field theory\ref{4,5}.
This conformal model is given by the following:
It contains four affine Kac-Moody currents $J^A$ of dimension one 
satisfying an $U(1)\times SU(2)$ KMA (of level $k$)
together with a set of four dimension-{$\hhalf$} fields $\psi^A$ satisfying 
the free fermion algebra:
$$ \eqalignno{J^0 (z) J^0 (\zeta) &= - (z-\zeta)^{-2} + \ldots\cr
J^0 (z) J^i (\zeta) &= O (z-\zeta)^0\cr
J^i (z) J^j (\zeta) &= - {k\over 2} (z-\zeta)^{-2} 
+ \epsilon_{ijk} J^k (\zeta)
(z-\zeta)^{-1} + \ldots\cr
\psi^A (z)\psi^B (\zeta)&= - (z-\zeta)^{-1} \delta^{AB} + \ldots\cr
\psi^A (z) J^B(\zeta)&= O (z-\zeta)^0\,.&(2.2)\cr}$$
The energy-momentum tensor from the Sugawara construction is:
$$\tilde L(z) = -\half J^0 J^0 - {1\over {k+2}} J^i J^i 
-\half\partial\psi^A \psi^A
+\delta {1\over 4 z^2}\,\eqno (2.3a)$$
where $\delta = 0,1$ in the NS or R sector respectively; and $A = (0,i)$.
The supercurrent is
$$ \tilde F(z) = \psi^0 J^0 + {{\sqrt 2}\over\sqrt{k+2}} \psi^i J^i
+ {{\sqrt 2}\over{6\sqrt{k+2}}} \epsilon_{ijk}\psi^i\psi^j\psi^k\,.
\eqno(2.3b)$$
The superconformal system (2.3) has $\tilde c = {6 (k+1)\over (k+2)}$.
In order to achieve the $c=6$ system corresponding to the fivebrane,
a shifted CFT is defined from (2.3): 
$$\eqalignno{\bar L(z) &= \tilde L(z) + \hhalf Q \partial J^0 (z) \cr
\bar F(z) &= \tilde F(z) -Q\partial\psi^0 (z)\,.&(2.4)\cr}$$
The superconformal system (2.3) has $c = \tilde c + 3 Q^2$.
We choose the background charge $Q^2 = {2\over{k+2}}$, so
that $c = 6$. Since we choose a real value for $Q$, this corresponds to the
strong coupling region of Liouville theory\ref{10,11}.

We can now construct 
the standard $N=4$ superconformal global algebra with generators 
$\bar L, \bar F^A,
\bar {\cal S}^i$ where $\bar {\cal S}^i$ generate a global
affine $SU(2)$ level $n$ algebra, and $c = 6n$.
The semi-wormhole representation 
for this algebra has $c=6$, i.e. $n=1$ and is given by
$$\eqalignno
{\bar L(z)&= -\half J^0 J^0 - {1\over {k+2}} J^i J^i 
-\half\partial\psi^A \psi^A
+\delta {1\over 4 z^2} +\half Q\partial J^0\cr
\bar F(z)&= \psi^0 J^0 + {{\sqrt 2}\over\sqrt{k+2}} \psi^i J^i
+ {{\sqrt 2}\over{6\sqrt{k+2}}}
\epsilon_{ijk}\psi^i\psi^j\psi^k - Q\partial\psi^0\cr
\bar F^i(z)&= \psi^i J^0 + {{\sqrt 2}\over\sqrt{k+2}} [ -\psi^0 J^i 
+ \epsilon_{ijk} J^j \psi^k - 
\half\epsilon_{ijk} \psi^0\psi^j\psi^k]
- Q\partial\psi^i\cr
\bar {\cal S}^i(z)&= 
\half (\psi^0\psi^i + \half \epsilon_{ij\ell}\psi^j\psi^\ell)
&(2.5)\cr}$$
for $\delta = 0,1$ for NS or R sector respectively. We set
$Q = -{\sqrt 2\over {\sqrt {k+2}}}$ and the currents in (2.5)
close on the $N=4$ superconformal algebra given in appendix A.
For $\bar L(z)$ hermitian, i.e. $\bar L_n^\dagger = \bar L_{-n}$, then 
$J_n^{0\dagger} = - J^0_{n}$ for $n\ne 0$, $J_0^{0\dagger} = -Q - J^0_0$,
$\psi_n^{A\dagger} = -\psi_{-n}^A$, and  $J_n^{i\dagger} = -J^i_{-n}$. 
Therefore the states
$J^0_{-n} |\psi\rangle$ for $n>0$ have positive norm:
$$|| J^0_{-n} |\psi\rangle ||^2 = \langle\psi |
(-J^0_n) J^0_{-n} |\psi\rangle =
n \langle \psi |\psi\rangle = n || |\psi\rangle ||^2 > 0$$ for
$||  |\psi\rangle ||^2 > 0$. A similar argument holds for
$\psi^A_{-n} |\psi\rangle$ and $J^i_{-n} |\psi\rangle$.
It follows from the above hermiticity conditions on $J_n^A,\psi_n^A$ 
that $\bar F_n^\dagger = \bar F_{-n}, \bar F_n^{i\dagger} = \bar F^i_{-n},$
and $\bar {\cal S}_n^{i\dagger} = - \bar {\cal S}^i_{-n}$. 

In the string models we consider, 
with the conformal field theory in (2.2) 
as a building block, we will use an $N=1$ superconformal ghost system,
so the $N=4$ algebra serves as a global worldsheet algebra, which has highest 
weight representations\ref{12}. Its $N=1$
subalgebra $\bar L, \bar F$ will be used as the matter (i.e. non-ghost)
contribution to the BRST charge operator.
We note that although $\bar {\cal S}^i(z)$ 
in (2.5) is a primary conformal field,
it is not the upper component of a primary superfield under the 
$N=1$ subalgebra.
In order to study vertex operators, we consider the primary
weight one-half superfield whose components are
$$V^i_L(z) = \psi^i \,;\qquad V^i_U (z) = Q
(\hhalf \epsilon_{ijk}\psi^j\psi^k + 
J^i )\equiv T^i\,,\eqno(2.6)$$
where
$$\eqalignno{\bar F(z) \psi^i(\zeta)&= (z-\zeta)^{-1}
T^i(\zeta)\cr
\bar F(z) T^i(\zeta) &= (z-\zeta)^{-2} \psi^i(\zeta)
+ (z-\zeta)^{-1}\partial\psi^i(\zeta)\,.&(2.7)\cr}$$ 
We define $T^i\equiv  V^i_U$ and (2.8)
forms a representation of a super Kac-Moody algebra
$$\eqalignno
{&T^i(z) T^j(\zeta) = - {\delta_{ij}\over (z-\zeta)^2} +  
{Q\epsilon_{ijk}T^k(\zeta)\over (z-\zeta)}&(2.8a)\cr
&T^i(z) \psi^j(\zeta) = {Q\epsilon_{ijk} \psi^k(\zeta)\over (z-\zeta)}
&(2.8b)\cr
&\psi^i(z) \psi^j(\zeta) = -{\delta_{ij}\over (z-\zeta)}\,.&(2.8c)\cr}$$
The level of the KMA $T^i$ is $k+2$. 
If we bosonize the $J^0$ current by
$J^0(z)=- \partial\phi(z)$, where
$\phi (z) \phi(\zeta) = - {\rm ln} (z-\zeta) + \ldots$ for $|z|>|\zeta|$,
then the conformal field $:e^{\beta\phi(z)}:$ with  
$:e^{\beta\phi(0)}: |0\rangle = |\beta\rangle$ 
is primary with respect to $\bar L(z)$
with conformal weight $h_Q = -\hhalf \beta (\beta + Q)$.
The operator product  
$$J^0(z) :e^{\beta\phi(\zeta)}: = \beta :e^{\beta\phi(\zeta)}: (z-\zeta)^{-1}
+\ldots \eqno(2.9)$$ implies $J^0_0 |\beta\rangle = \beta |\beta\rangle$.
The anomalous product
$$\bar L(z) J^0(\zeta) = Q (z-\zeta)^{-3} + J^0 (\zeta) (z-\zeta)^{-2}
+ \partial J^0 (\zeta) (z-\zeta)^{-1} + \ldots \,\eqno(2.10)$$ implies 
$[L_n, J_m] = -m J_{n+m} + {Q\over 2} n(n+1)\delta_{n,-m}$.
For a CFT with background charge, 
considering the moments of the current $J^0(z)$ and requiring 
$\bar L_n^\dagger = \bar L_{-n}$, we derive  
$$J_0^{0\dagger} = - [ \bar L_{-1}, J^0_1]^\dagger =
- [ \bar L_{1}, J^0_{-1}] = - J_0^0 - Q\,.\eqno (2.11)$$
The allowed values for $\beta$ such that $h_Q$ is
real are given below (4.11). The case we consider in this paper is
real $Q$ and real $h_Q$.  

We now discuss in detail an inner product for the Liouville states with
real $\beta$, i.e. the discrete states, by describing 
an inner product for the vector space of states whose
elements, i.e. vectors,  
are linear combinations of the states $|\beta\rangle$, including
the unique element ${\bf 0}$ defined by $|\beta\rangle - |\beta\rangle = 
{\bf 0}$. We note $|0\rangle$ is different from ${\bf 0}$.
The choice of an inner product in a vector space is by no means 
uniquely determined by the vector space. The choice of an inner product
is a free choice, which leads us to consider the question of exactly what
is chosen, i.e. what is an inner product.\ref{13} 

An inner product on a real vector space is a real-valued function
which assigns to any ordered pair of vectors such as $|\beta\rangle$,
$|\beta'\rangle$ 
a real number, denoted by 
$(|\beta'\rangle , |\beta\rangle )$ and satisfying
i) $(|\beta'\rangle , |\beta\rangle ) =
(|\beta\rangle , |\beta'\rangle )$, i.e. the inner product is symmetric, 
ii) $(|\beta'\rangle , a |\beta\rangle ) 
= a (|\beta'\rangle , |\beta\rangle )$, for any scalar $a$,
$\,$iii) $( |\beta'\rangle, [|\beta_1\rangle + |\beta_2\rangle ]) =
( |\beta'\rangle, |\beta_1\rangle ) + ( |\beta'\rangle, |\beta_2\rangle )$,
$\,$ iv)\nobreak$(|\beta\rangle , |\beta\rangle ) > 0$, 
and v) $({\bf 0}, {\bf 0} ) = 0,\,\, ({\bf 0}, |\beta\rangle ) =
(|\beta\rangle , {\bf 0}),\, ({\bf 0} , a |\beta\rangle )
= a ({\bf 0} , |\beta\rangle )$, for any scalar $a$,
and $( {\bf 0}, [|\beta_1\rangle + |\beta_2\rangle] ) =
({\bf 0}, |\beta_1\rangle ) + ( {\bf 0}, |\beta_2\rangle )$.

We now choose a particular inner product for the Liouville states
$|\beta\rangle$. We define
the inner product $(|\beta'\rangle , |\beta\rangle )$ such that
$$(|\beta'\rangle , J_0^0|\beta\rangle ) =
(J_0^{0\dagger} |\beta'\rangle , |\beta\rangle )\eqno(2.12)$$ where
the lefthand side of (2.12) is 
$$(|\beta'\rangle , J_0^0|\beta\rangle ) 
= \beta ( |\beta'\rangle,|\beta\rangle )\,.\eqno(2.13)$$
The righthand side of (2.12) is
$$\eqalignno {(J_0^{0\dagger} |\beta'\rangle , |\beta\rangle ) &=
( \beta' |\beta'\rangle, |\beta\rangle )\,.&(2.14)\cr}$$
The choice of inner product is defined by (2.14).
It is equivalent to defining the adjoint of the state $|\beta\rangle$ to
be $\langle -\beta - Q|$ in the sense that
$$\eqalignno{J^0_0 |\beta\rangle &= \beta |\beta\rangle&(2.15a)\cr 
[ J^0_0 |\beta\rangle ]^\dagger 
&= \beta \langle -\beta - Q |\,.\cr
&&(2.15b)\cr}$$
So that in (2.12) the righthand side 
$(J_0^{0\dagger} |\beta'\rangle , |\beta\rangle )$ means  
$( [\,\langle -\beta' - Q | J_0^0\,]^\dagger\,, |\beta\rangle\,)$.
Here (2.15b) for $\beta = \beta'$, which is 
$\langle -\beta' -Q | J_0^{0\dagger}\,
= \beta' \langle -\beta' -Q |$, is the definition
used in deriving (2.14), i.e. our choice for inner product is 
$$(|\beta'\rangle, |\beta\rangle) \equiv \langle -\beta' - Q |\beta\rangle
\,.\eqno(2.16)$$
This choice satisfies the conditions i)--v) listed above. 
{}From (2.12)-(2.14), we see that 
$(|\beta'\rangle, |\beta\rangle)$ is non-vanishing only for
$\beta = \beta'$, and we normalize
$(|\beta\rangle, |\beta\rangle) = 1$ . In the bra, ket notation above, 
this is 
$$\langle -\beta -Q | \beta \rangle = 1\,\eqno(2.17)$$ 
We also have
$$ \langle 0 | e^{-Q\phi (z)} | 0\rangle =1\,.\eqno(2.18)$$
With the inner product defined by (2.14), the discrete Liouville states
have positive norm: 
$$ || |\beta\rangle ||^2 \equiv 
(|\beta\rangle, |\beta\rangle) = 1\,.\eqno(2.19)$$
In sect. 5, we consider the holomorphic
$N=1$ $SU(2)$ super Kac-Moody
algebra mixing with this CFT and show how these discrete states allow for
massless Ramond states (4d spacetime fermions) {\it which carry non zero
$SU(2)$ charge} on a given side of a closed string theory.  

\vskip15pt
\centerline{\bf3. The string model}
\vskip5pt
We consider compactification of the Type II superstring with  
four-dimensional Lorentz invariance.
There is a $c=9$, $N=1$ internal (local) superconformal
algebra (SCA) associated with any such compactification. 
For theories with $d=4$, $N=2$ spacetime supersymmetry,
the spacetime supersymmetry implies that
this internal SCA splits into a $c=6$ piece with $N=4$ 
superconformal (global) symmetry and a $c=3$ piece with 
$N=2$ superconformal (global) symmetry.\ref{14,15}
Here we use {\it local} to
mean the BRST ghost system also satifies this algebra. 

We consider the $M^{(4)} \otimes T^{(2)} \otimes W_k^{(4)}$ model,
where $W_k^{(4)}$ is the $c=6$ fivebrane
CFT given in (2.4), and $M^{(4)} \otimes T^{(2)}$ is the  
flat four-dimensional non-compact flat Minkowski spacetime
times a two-dimensional compact flat torus, given in terms of 
six free superfields $M = (0\le \mu\le 3; 4\le m \le 5 )$,
$$ \eqalignno{ a^M (z) a^N (\zeta)
&= \eta^{MN} (z-\zeta)^{-2} + \ldots\cr
\psi^M (z) \psi^N (\zeta) &= \eta^{MN} (z-\zeta)^{-1} + \ldots\cr
\psi^M (z)  a^N (\zeta) &= 0&(3.1)\cr}$$
where one of the free superfields is chosen to be timelike, and the
others to be Euclidean (spacelike); $\eta^{MN} = (-1, 1, 1, 1, 1, 1)$
and $a^M(z)\equiv i\partial X^M (z)$ where
$$\eqalignno{ X^M(z) X^N(\zeta) &= - \eta^{MN} 
\ln (z-\zeta) + : X^M(z) X^N(\zeta) :\cr
\psi^M(z) \psi^N(\zeta) &= \eta^{MN} (z-\zeta)^{-1} +\dots\cr
\psi^M(z) a^N(\zeta) &= O(z-\zeta)^0\,.&(3.2)\cr}$$
Then 
$$\eqalignno{L(z) &= \half :a^M a_M : + \half\partial\psi^M\,\psi_M
-\half J^0 J^0 - \hhalf Q^2 J^i J^i -\half\partial\psi^A \psi^A
+\delta {5\over 8 z^2} +\half Q\partial J^0\cr
F(z)&= a^M\psi_M +  \psi^0 J^0 - Q \psi^i J^i
- {Q\over 6} 
\epsilon_{ijk}\psi^i\psi^j\psi^k - Q\partial\psi^0&(3.3)\cr}$$
for $\delta = 0,1$ for NS or R sector respectively, where 
$Q = - {\sqrt 2\over\sqrt {k+2}}$ and $c=15$. 

\vskip15pt
\centerline{\bf 4. Unitary representations of the $N=4$ superconformal
algebra}
\vskip10pt
\leftline{NEVEU-SCHWARZ STATES}

We analyze the BRST properties\ref{16} and gauge symmetry of the following 
four-dimensional vertex operators, given here in covariant
form where $\mu$ labels 
spacetime, $0\le\mu\le 3$, and the metric is $\eta_{\mu\nu} = {\rm diag}
(-1, 1, 1, 1)$. 
The weight one primary conformal fields for 
Neveu-Schwarz states with one oscillator 
are given in the superconformal ghost number $q=-1$ picture by
$$\eqalignno{ V_{-1}(k,z,\epsilon) 
&= \epsilon\dt\psi (z) e^{ik_\mu X^\mu (z)} 
e^{\beta\phi(z)} V_{jm} (z) e^{-\varphi_{\rm gh}(z)}&(4.1a)\cr
V^i_{-1}(k,z) &= \psi^i (z) e^{ik_\mu X^\mu (z)}    
e^{\beta\phi(z)} V_{jm} (z) e^{-\varphi_{\rm gh}(z)}&(4.1b)\cr
V^0_{-1}(k,z) &= \psi^0 (z) e^{ik_\mu X^\mu (z)}              
e^{\beta\phi(z)} V_{jm} (z) e^{-\varphi_{\rm gh}(z)}&(4.1c)\cr
V^n_{-1}(k,z) &= \psi^n (z) e^{ik_\mu X^\mu (z)}    
e^{\beta\phi(z)} V_{jm} (z) e^{-\varphi_{\rm gh}(z)}\,,&(4.1d)\cr}$$
where the weight one condition implies
$k_\mu k^\mu -\beta (\beta + Q) + Q^2 j(j+1) = 0$ for each field. 
Also $1\le n\le 6$ where the $\psi^n$ form a $c=3$ superconformal system
corresponding to two supercoordinates $\psi^\ell + \theta a^\ell,
\ell=1,2$ compactified on $T^2$. 
The fields $V_{jm}(z)$, which are primary under the superconformal 
algebra as well as the level $k$ SU(2) affine algebra, form $SU(2)$ multiplets 
with $m = ( -j, -j+1, \dots, j)$. Here $V_{jm}(z)$ corresponds
to a unitary highest weight representation of affine $SU(2)_k$, so that 
$j = 0, \half, 1, \ldots, {k\over 2}$.
$$\eqalignno{L(z) V_{jm}(\zeta)&= (z-\zeta)^{-2} 
\,{\textstyle{j(j+1)\over {k+2}}}\, V_{jm}(\zeta) 
+ (z-\zeta)^{-1}\partial_\zeta V_{j,m}(\zeta)\cr
J^i(z) V_{jm}(\zeta)&= (z-\zeta)^{-1} f_{imm'} V_{jm}(\zeta)\,,&(4.2)\cr}$$
where $f_{imm'}= t^i_{mm'}$ is the $SU(2)$ representation of 
$V_{jm}(\zeta)$.
Without the ghosts, states in the Neveu-Schwarz system form superconformal
fields $V(z,\theta) = V_q(z) + \theta V_{q+1}(z)$ with upper and 
lower components related by 
$$\eqalignno{G(z) V_q(\zeta) &= (z-\zeta)^{-1} V_{q+1}(\zeta)\cr
G(z) V_{q+1} (\zeta) &= (z-\zeta)^{-2} 2h_q V_q(\zeta) +
(z-\zeta)^{-1}\partial V_q(\zeta)\,.&(4.3)\cr}$$

The BRST current for the $c=15$ superVirasoro currents $L, F$ is
the usual expression
$$\eqalignno{Q(z) &= Q_0(z) + Q_1(z) +Q_2(z)\cr
Q_0(z) &= c L - \nox c b\partial c\nox
+ {\textstyle{3\over 2}}{\partial^2 c}
+ c L^{\beta\gamma}
+ \partial({\textstyle{3\over 4}}\nox\gamma \beta \nox )\cr
Q_1(z) &= -\hhalf e^{(\chi_{\rm gh} + \varphi_{\rm gh})} F\cr
Q_2(z) &= -{\textstyle{1\over 4}}\gamma b\gamma \,.&(4.4)\cr}$$ 
BRST invariance holds for the vertices (4.1) with 
the following restrictions. For 
$$F(z) = a_\mu(z) \psi^\mu(z) + \bar F(z) 
+ a_\ell (z)\psi^\ell (z)\,\eqno(4.5)$$ and
$$\bar F(z) = \psi^0 J^0 - Q \psi^i J^i 
-Q  {1\over 6}
\epsilon_{ijk}\psi^i\psi^j\psi^k - Q\partial\psi^0 \eqno(4.6)$$
where the background charge is $Q = - {\scriptstyle{\sqrt{2\over {2+k}}}}$,
we have
$$\eqalignno{F(z) V_{-1}(k,\zeta,\epsilon) &=
(z-\zeta)^{-1} [ (k\dt\psi(\zeta)\epsilon\dt\psi(\zeta)
+ \epsilon\dt a(\zeta) + \beta\psi^0(\zeta)\epsilon\dt\psi(\zeta) )
V_{jm}(\zeta)\cr 
& \hskip 50pt - Q 
\psi^i(\zeta)\epsilon\dt\psi(\zeta) 
f_{imm'} V_{jm'}(\zeta)] e^{ik_\mu X^\mu (\zeta)}
e^{\beta\phi(\zeta)}&(4.7a)\cr
F(z) V^i_{-1}(k,\zeta) &= 
(z-\zeta)^{-2} [ Q f_{imm'} V_{jm'}(\zeta)]
e^{ik_\mu X^\mu (\zeta)}
e^{\beta\phi(\zeta)}\cr
&+ (z-\zeta)^{-1} [ (k\dt\psi(\zeta)\psi^i(\zeta)
+ \beta\psi^0(\zeta)\psi^i(\zeta)
+ Q \hhalf\epsilon_{ikk'}
\psi^k(\zeta)\psi^{k'}(\zeta) ) V_{jm}(\zeta)\cr
&\hskip 50pt + Q ( :J^i(\zeta) V_{jm}(\zeta): 
- \psi^k(\zeta)\psi^i(\zeta) f_{kmm'} V_{jm'}(\zeta) ) ]
e^{ik_\mu X^\mu (\zeta)}                                                  
e^{\beta\phi(\zeta)}\cr
&&(4.7b)\cr            
F(z) V^0_{-1}(k,\zeta) &=
(z-\zeta)^{-2} [ (-\beta - Q) V_{jm}(\zeta)]   
e^{ik_\mu X^\mu (\zeta)}   
e^{\beta\phi(\zeta)}\cr
&+ (z-\zeta)^{-1} [ ( k\dt\psi(\zeta)\psi^0(\zeta)e^{\beta\phi(\zeta)}
+ :\partial\phi(\zeta) e^{\beta\phi(\zeta)}: )
V_{jm}(\zeta)\cr
& \hskip 50pt - Q 
\psi^k(\zeta)\psi^0(\zeta)                      
f_{kmm'} V_{jm'}(\zeta) e^{\beta\phi(\zeta)}] 
e^{ik_\mu X^\mu (\zeta)}&(4.7c)\cr
F(z) V^n_{-1}(k,\zeta) &= (z-\zeta)^{-1} [ (k\dt\psi(\zeta)\psi^n(\zeta)
+ \beta\psi^0(\zeta)\psi^n(\zeta) 
-\hhalf \epsilon_{nn'n''} \psi^{n'}(\zeta)\psi^{n''}(\zeta))
V_{jm}(\zeta)\cr
& \hskip 50pt - Q
\psi^k(\zeta)\psi^n(\zeta)
f_{kmm'} V_{jm'}(\zeta)] e^{ik_\mu X^\mu (\zeta)}
e^{\beta\phi(\zeta)}\,.&(4.7d)\cr}$$
In (4.7), the four expressions $V_{-1}$ are used without the ghost field.
BRST invariance of the vertex operators is 
${1\over {2\pi i}}\oint_\zeta dz Q(z) V_{-1}(\zeta) = 0$ which thus 
requires
$V_{jm}(z) = V_{00}(z)$ in (4.1b) and $\beta = -Q$ in 
(4.1c). With these restrictions, we have
$F_s V_{-1}(k,0) |0\rangle =0$ for $s\ge\hhalf$; so that 
$V_{-1}(k,0)|0\rangle$ satisfy the physical state conditions. 
We note that
$[L_{-1}, V_{jm} (\zeta) ] = -Q^2 f_{imm'} :J^i(\zeta)V_{jm}(\zeta):
=\partial V_{jm}(\zeta)$. Therefore $\partial V_{00} = 0$, so 
$V_{00}(\zeta)$ is a constant, and in what follows we can set it equal to
one.
We then define the upper components of (4.1), which satisfy (4.3) as
$$\eqalignno{ V_{0}(k,\zeta,\epsilon)
&= [(k\dt\psi(\zeta)\epsilon\dt\psi(\zeta)
+ \epsilon\dt a(\zeta) + \beta\psi^0(\zeta)\epsilon\dt\psi(\zeta) )
V_{jm}(\zeta)\cr
& \hskip 5pt- Q
\psi^i(\zeta)\epsilon\dt\psi(\zeta)
f_{imm'} V_{jm'}(\zeta)] e^{ik_\mu X^\mu (\zeta)}
e^{\beta\phi(\zeta)}&(4.8a)\cr
V^i_{0}(k,\zeta) &= 
[ k\dt\psi(\zeta)\psi^i(\zeta)
+ \beta\psi^0(\zeta)\psi^i(\zeta)\cr
& \hskip 5pt+ Q ( J^i(\zeta) + \hhalf\epsilon_{ikk'}
\psi^k(\zeta)\psi^{k'}(\zeta) )  ] 
e^{ik_\mu X^\mu (\zeta)}
e^{\beta\phi(\zeta)}&(4.8b)\cr
V^0_{0}(k,z) &= [ ( k\dt\psi(\zeta)\psi^0(\zeta)e^{-Q\phi(\zeta)}
+ :\partial\phi(\zeta) e^{-Q\phi(\zeta)}: )
V_{jm}(\zeta)\cr
& \hskip 5pt - Q
\psi^k(\zeta)\psi^0(\zeta)
f_{kmm'} V_{jm'}(\zeta) e^{-Q\phi(\zeta)}] 
e^{ik_\mu X^\mu (\zeta)}&(4.8c)\cr 
V^n_{0}(k,z) &= 
[ (k\dt\psi(\zeta)\psi^n(\zeta)
+ \beta\psi^0(\zeta)\psi^n(\zeta)
-\hhalf \epsilon_{nn'n''} \psi^{n'}(\zeta)\psi^{n''}(\zeta))
V_{jm}(\zeta)\cr
& \hskip 5pt - Q
\psi^k(\zeta)\psi^n(\zeta)
f_{kmm'} V_{jm'}(\zeta)] e^{ik_\mu X^\mu (\zeta)}
e^{\beta\phi(\zeta)}\,.&(4.8d)\cr}$$
{}From (4.1) we see that the states $V_{-1}(k,0)|0\rangle$ are
tensor products of states in the three conformal field theories: 
$({\rm\, spacetime\,cft}\, 
c=6)\otimes ({\rm fivebrane\, cft}\,c=6)\otimes (T^2 {\rm cft}\, c=3)$.
In order to specify the quantum numbers $\beta$ and $j$ in (4.1),
we consider the conditions for highest weight representations\ref{12} of the 
global $N=4$ superconformal algebra constructed from the 
$c=6$ fivebrane CFT as given in (2.5) and Appendix A.  
In the Neveu-Schwarz sector, the highest weight states of the 
representations
are annihilated by 
$\bar L_n, \bar F_r, \bar F^i_r, \bar {\cal S}^i_n, \bar {\cal S}^+_0$
for $n\ge 1, r\ge\hhalf$ and
are labelled by
$$\bar L_0 |h,\ell\rangle = h |h,\ell\rangle\,,\qquad
i \bar {\cal S}_0^3 |h,\ell\rangle = \ell |h,\ell\rangle \,,\eqno(4.9)$$
where $\ell = 0$ and $h>0$ for massive representations; and
$h=\ell$ with $\ell = 0,\hhalf$ for massless representations\ref{12}.
For massless representations, the highest weight states $|h,\ell\rangle$
are also annihilated by $\bar{\cal G}^1_{-\half}$ and ${\cal G}^2_{-\half}$  
with ${\cal G}^a$ given in Appendix A:
$$\bar{\cal G}^1_{-\half} |h,\ell\rangle = 0\,;\qquad
{\cal G}^2_{-\half} |h,\ell\rangle = 0\,.\eqno(4.10)$$
{\it For states with no $(N=4)$ oscillators as in }(4.1a,d), we have
$$\bar L_0 \,e^{\beta\phi(0)} V_{jm}(0) |0\rangle
= [ -\hhalf\beta(\beta + Q) +\hhalf Q^2 j (j+1) ] 
e^{\beta\phi(0)} V_{jm}(0) |0\rangle\,. \eqno(4.11)$$
For $h_Q \equiv -\hhalf\beta(\beta + Q)$ to be
real, either $\beta = -\hhalf Q + iy$ with real $y$ so that 
$h_Q = {\textstyle{1\over 8}}Q^2 + \hhalf y^2$ is positive;
or $\beta$ is real. If $\beta$ is real, the highest weight conditions
for massless representations
require it to take on discrete values, a feature similar to the discrete
states in $c=1$ matter coupled to $2d$ quantum gravity, 
i.e. the Liouville field.\ref{17,11}
Also
$$i\bar {\cal S}_0^3 \,e^{\beta\phi(0)} V_{jm}(0) |0\rangle =  
{\textstyle{i\over 2}}\sum_s (\psi^0_{-s}\psi^i_s 
+ \half \epsilon_{ij\ell}\psi^j_{-s}\psi^\ell_s) 
e^{\beta\phi(0)} V_{jm}(0) |0\rangle = 0\,,\eqno(4.12)$$
so that $e^{(-{Q\over 2} + iy)\phi(0)} V_{jm}(0) |0\rangle$
correspond to $N=4$ highest weight massive states:
 $h>0$, $\ell=0$. The states 
$e^{\beta\phi(0)} V_{jm}(0) |0\rangle$ for $\beta = Q j, -Q (j+1)$ 
have $h=\ell = 0$, and from (4.10) 
correspond to $N=4$ highest weight massless states for $\beta = Qj =0$.

\noindent {\it For states with one $(N=4)$ oscillator as in }(4.1b,c), 
we have
$$\eqalignno{\bar L_0 \,\psi^{(i,0)i}(0) 
e^{\beta\phi(0)} V_{jm}(0) |0\rangle
&= [ \hhalf -\hhalf\beta(\beta + Q) +\hhalf Q^2 j (j+1) ]
\,\psi^{(i,0)}(0) e^{\beta\phi(0)} V_{jm}(0) |0\rangle\,.\cr
&&(4.13)\cr}$$
Also
$$\eqalignno{i\bar {\cal S}_0^3 \, (\psi^1(0) \pm i\psi^2(0)) 
e^{\beta\phi(0)} V_{jm}(0) |0\rangle &=
\pm\hhalf\, 
(\psi^1(0) \pm i\psi^2(0)) e^{\beta\phi(0)} V_{jm}(0) |0\rangle\cr
i\bar {\cal S}_0^3 \,(\psi^3(0) \mp i\psi^0(0)) 
e^{\beta\phi(0)} V_{jm}(0) |0\rangle &=
\pm\hhalf\,(\psi^3(0) \mp i\psi^0(0))  
e^{\beta\phi(0)} V_{jm}(0) |0\rangle \,,&(4.14)\cr}$$ 
and we find that 
$$(\psi^1(0) + i\psi^2(0) ) e^{-Q\phi(0)} |0\rangle\,;\qquad 
(\psi^3(0) - i\psi^0(0) ) e^{-Q\phi(0)} |0\rangle\,\eqno(4.15)$$
satisfy the $N=4$ 
highest weight massless state conditions with $h=\ell=\hhalf$.
The states
$\psi^i(0)\,e^{(-{Q\over 2} + iy)\phi(0)} |0\rangle$
and $\psi^0(0)\,e^{(-{Q\over 2} + iy)\phi(0)} |0\rangle$ 
are not $N=4$ highest weight massive states since both $F_s$, $F^i_s$ 
acting on
either of them is not zero, for all $s\ge\hhalf$.
BRST invariant vertex operators comprised of 
worldsheet $N=4$ massive representations are
$$ \eqalignno{ V_{-1}(k,z,\epsilon)
&= \epsilon\dt\psi (z) e^{ik_\mu X^\mu (z)}
e^{(-{Q\over 2} + iy) \phi(z)} 
V_{jm} (z) e^{-\varphi_{\rm gh}(z)}&(4.16a)\cr 
V^n_{-1}(k,z) &= \psi^n (z) e^{ik_\mu X^\mu (z)}
e^{(-{Q\over 2} + iy) \phi(z)} 
V_{jm} (z) e^{-\varphi_{\rm gh}(z)}\,,&(4.16b)\cr}$$
with $k_\mu k^\mu + {Q^2\over 4} + y^2 + Q^2 j (j+1)  = 0$.

BRST invariant vertex operators comprised of
worldsheet $N=4$ massless representations are      
$$\eqalignno{ V_{-1}(k,z,\epsilon) &=
\epsilon\dt\psi (z) e^{ik_\mu X^\mu (z)}
e^{-\varphi_{\rm gh}(z)}&(4.17a)\cr
V^i_{-1}(k,z) &= \psi^i (z) e^{ik_\mu X^\mu (z)}
e^{-Q\phi(z)} e^{-\varphi_{\rm gh}(z)}&(4.17b)\cr
V^0_{-1}(k,z) &= \psi^0 (z) e^{ik_\mu X^\mu (z)}
e^{-Q\phi(z)} e^{-\varphi_{\rm gh}(z)}&(4.17c)\cr
V^n_{-1}(k,z) &= \psi^n (z) e^{ik_\mu X^\mu (z)}
e^{-\varphi_{\rm gh}(z)}\,&(4.17d)\cr}$$ 
with
$k_\mu k^\mu = 0$. 
It follows that states corresponding to the vertex operators 
in (4.16,17): $|\psi\rangle = V_{-1}(k,0) |0\rangle$ are 
representations of the direct product worldsheet algebras and 
satisfy the physical state conditions:
$$ F_s |\psi\rangle = 0, \,{\rm for}\,s\ge\hhalf;\quad
L_n |\psi\rangle = 0, \,{\rm for}\,n\ge 1;\quad
L_0 |\psi\rangle = \hhalf |\psi\rangle\,.\eqno(4.18)$$
{}From (4.17b,c) and (4.14)  we see that 
$V_{-1}^i(k,0) |0\rangle$ and $V_{-1}^0(k,0) |0\rangle$
are doublets under the $SU(2)$ algebra $\bar {\cal S}^i_0$ whose
affine algebra $\bar {\cal S}^i(z)$ are 
the level one $SU(2)$ generators of the
$N=4$, $c=6$ global superconformal algebra (2.5).

{}From (2.8b) however, we note that the states
$$\tilde V_{-1}^i(k,0) |0\rangle = \psi^i (0) e^{ik_\mu X^\mu (0)}
e^{-\varphi_{\rm gh}(0)}|0\rangle\,,\qquad 
V_{-1}^0(k,0) |0\rangle\eqno(4.19a)$$
are highest weight states of the $N=1$ subalgebra, 
and are in the adjoint and singlet representations of the $SU(2)$
algebra $T^i_0$, whose current $T^i(z)$ is the upper component of 
$\psi^i(z)$. Therefore
$\tilde V_{-1}^i(k,0) |0\rangle$ 
correspond to gauge bosons of $SU(2)$, i.e.
their three-point  correlation functions would have the 
standard non-abelian three-point coupling.
$$\eqalignno{
i Q [ J^i(z) +\hhalf\epsilon_{ikp}\psi^k(z)\psi^p(z)] 
\tilde V_{-1}^j(k,\zeta)
&= (z-\zeta)^{-1}  
Q i\epsilon_{ijk} \tilde V_{-1}^k(k,\zeta)\,.&(4.19b)\cr}$$

\hskip12pt

\leftline{RAMOND STATES}
The Ramond ground states of the $N=4$ superconformal algebra are expressed 
in terms of spin fields $S_A$, $S_{\dot A}$.
Instead of (4.3), the $N=4$ Ramond fields 
satisfy
$$\eqalignno{\bar F(z) \,S_A(\zeta) e^{\beta\phi(\zeta)}
V_{jm}(\zeta) &= (z-\zeta)^{-{3\over 2}} 
[(\beta + Q ){\textstyle{1\over \sqrt 2}} \gamma_A^{\hskip3pt 0\dot B}
S_{\dot B}(\zeta) V_{jm}(\zeta) \cr
&\hskip 50pt- Q f_{imm'}  {\textstyle{1\over \sqrt 2}} 
\gamma_A^{\hskip3pt i\,\dot B} 
S_{\dot B}(\zeta) V_{jm'}(\zeta)] \,e^{\beta\phi(\zeta)}&(4.20a)\cr
&\hskip 5pt+ O (z-\zeta)^{-{1\over 2}}\cr
\bar F(z) \, S_{\dot A} (\zeta) e^{\beta\phi(\zeta)}
V_{jm}(\zeta)
&= (z-\zeta)^{-{3\over 2}}
[\beta {\textstyle{1\over \sqrt 2}} \gamma_{\dot A}^{\hskip3pt 0 B}
S_{B}(\zeta) V_{jm}(\zeta) \cr
&\hskip 50pt- Q f_{imm'}  {\textstyle{1\over \sqrt 2}} 
\gamma_{\dot A}^{\hskip3pt i\, B}
S_{B}(\zeta) V_{jm'}(\zeta)] \,e^{\beta\phi(\zeta)}\cr
&\hskip5pt+ O (z-\zeta)^{-{1\over 2}}\,.&(4.20b)\cr}$$
Here we choose a Weyl representation for the four-dimensional $\gamma$ matrix 
algebra with $\{\gamma^m,\gamma^n\} = - 2 \delta^{mn}$ given by
$$\gamma^0 =
\left(\matrix{0&(-\sigma^0)^A_{\hskip3pt\dot B}\cr
(\sigma^0)^{\dot A}_{\hskip3pt B}&0\cr}\right)\,; \qquad
\gamma^j = i 
\left(\matrix{0&(\sigma^j)^A_{\hskip3pt\dot B}\cr
(\sigma^j)^{\dot A}_{\hskip3pt B}&0\cr}\right)\,\eqno(4.21a)$$
where $\sigma^0\equiv\left(\matrix{1&0\cr 0&1\cr}\right)$ and
$\sigma^i$ are the Pauli matrices.
The charge conjugation matrices
$$C = \left(\matrix{(i\sigma^2)^{AB}&0\cr
0&(i\sigma^2)^{\dot A\dot B}\cr}\right)\,; \qquad
C^{-1} = \left(\matrix{(-i\sigma^2)_{AB}&0\cr
0&(-i\sigma^2)_{\dot A\dot B}\cr}\right)\,\eqno(4.21b)$$
raise and lower indices:
$C^{-1}_{AD} (\gamma^m)^D_{\hskip3pt\dot B} =
(\gamma^m)_{A\dot B}$ and
$C^{\dot B\dot D} (\gamma^m)^A_{\hskip3pt\dot D} =
(\gamma^m)^{A\dot B}$, etc. 
In deriving (4.20), we use the operator product expansion
$$\eqalignno{\psi^{(0,i)}(z) S_A(\zeta) &=
(z-\zeta)^{-\half} {\textstyle{1\over \sqrt 2}} 
\gamma_A^{\hskip3pt (0,i)\dot B}
S_{\dot B}(\zeta) +\dots\cr
\psi^{(0,i)}(z) S_{\dot A}(\zeta) &=
(z-\zeta)^{-\half} {\textstyle{1\over 
\sqrt 2}} \gamma_{\dot A}^{\hskip3pt(0,i)B}
S_{B}(\zeta) +\dots\,\cr
\psi^i(z)\psi^j(z) S_A (\zeta)&=
(z-\zeta)^{-1} {\textstyle{1\over
2}} (\gamma^i\gamma^j)_A^{\hskip3pt B}
S_{B}(\zeta) +\dots\,,{\rm etc.}&(4.22)\cr}$$
and use that in this representation 
${\textstyle{1\over 6}}\epsilon_{ijk} 
(\gamma^i\gamma^j\gamma^k)_{A}^{\hskip 3pt \dot B} 
= - \gamma_A^{\hskip3pt 0\dot B}$; and 
${\textstyle{1\over 6}}\epsilon_{ijk}     
(\gamma^i\gamma^j\gamma^k)_{\dot A}^{\hskip 3pt B}               
= \gamma_{\dot A}^{\hskip3pt 0 B}$. 

In the Ramond sector, the highest weight states of the 
$N=4$ representations are annihilated by
$\bar L_n, \bar F_n, \bar F^i_n, \bar {\cal S}^i_n, \bar {\cal S}^+_0, 
{\cal G}^2_0,
\bar{\cal G}^1_0$   for $n\ge 1,$ and are labelled by
$$\bar L_0 |h,\ell\rangle = h |h,\ell\rangle\,\qquad
i \bar {\cal S}_0^3 |h,\ell\rangle = \ell |h,\ell\rangle \,,\eqno(4.23)$$
where $\ell = \hhalf$ 
and $h>{\textstyle{1\over 4}}$ for massive representations; 
and $h={\textstyle{1\over 4}}$ 
with $\ell = 0,\hhalf$ for massless representations\ref{12}.
For massless representations, the highest weight states $|h,\ell\rangle$
are also destroyed by ${\cal G}^1_0$ and $\bar{\cal G}^2_0$, so that
$$\bar F_0 |h,\ell\rangle = 0\,;\qquad
\bar F^i_0 |h,\ell\rangle = 0\,.\eqno(4.24)$$
{}From (4.22), we find the global level one $SU(2)$ current acts 
on the spin field states as 
$$i\bar{\cal S}_0^3 \,S_1(0)|0\rangle = -\hhalf \,S_1(0)|0\rangle\,,\quad
i\bar{\cal S}_0^3 \,S_2(0)|0\rangle = \hhalf \,S_2(0)|0\rangle\,;\quad
i\bar{\cal S}_0^3 \,S_{\dot A}(0)|0\rangle = 0\,\eqno(4.25)$$
and from (2.5) and (4.20,22), we have that 
in the Ramond sector,
the highest weight $N=4$ massive state with $h>{\textstyle{1\over 4}}$,
$\ell = \hhalf$ is $S_2(0) e^{(-{Q\over 2}+iy)\phi(0))} V_{00}(0)|0\rangle$. 

The highest weight $N=4$ massless Ramond states 
with $h={\textstyle{1\over 4}}$ are, for $\ell = \hhalf$,
$S_A(0) e^{-Q\phi(0))} |0\rangle$; and 
$S_{\dot A}(0)|0\rangle$ for $\ell = 0$.

It is convenient to bosonize the free fermions in terms of two
bosons, $H^\pm (z)$ as
$$\bar {\cal S}^3 = {1\over\sqrt 2}\partial H^+,\quad
\bar {\cal S}^\pm = e^{\pm i\sqrt 2 H^+};\qquad
\tilde {\cal S}^3 = {1\over\sqrt 2}\partial H^-,\quad
\tilde {\cal S}^\pm = e^{\pm i\sqrt 2 H^-}\eqno(4.26)$$
where
$$\bar {\cal S}^i = 
\hhalf (\psi^0\psi^i + \half \epsilon_{ij\ell}\psi^j\psi^\ell)\,;\quad
\tilde {\cal S}^i = 
\hhalf (-\psi^0\psi^i + \half \epsilon_{ij\ell}\psi^j\psi^\ell)\eqno(4.27)$$ 
are two commuting level one $SU(2)$ currents. Since
$$i\bar {\cal S}^3(z) e^{\pm {i\over\sqrt 2} H^+(\zeta)}
=\pm\hhalf (z-\zeta)^{-1} e^{\pm {i\over\sqrt 2} H^+(\zeta)}\,;
\qquad i\bar {\cal S}^3(z) e^{\pm {i\over\sqrt 2} H^-(\zeta)} = 0\,,
\eqno(4.28)$$
we identify 
$$S_A(z) =  e^{\mp {i\over\sqrt 2} H^+(z)}\,\qquad
S_{\dot A}(z) =  e^{\mp{i\over\sqrt 2} H^-(z)}\,.\eqno(4.29)$$
In general, modular invariance requires certain projections of the 
spin fields which restrict the physical spectrum.
In the direct product of the three conformal field 
theories: 
( spacetime cft $c=6$) 
$\otimes\, ({\rm fivebrane\, cft}\,c=6)\otimes (T^2 {\rm cft}\, c=3)$,
we consider the BRST invariant 
weight one primary conformal fields for Ramond ground states 
in the superconformal ghost number $q=-\hhalf$ picture comprised of
$N=4$ massless highest weight representations given by  
$$\eqalignno{ V^{(1)}_{-\half}(k,z)
&= u^{1\alpha} (k) S_\alpha (z) e^{ik_\mu X^\mu (z)}
S_A(z) e^{{-Q}\phi(z)} S_n(z) 
e^{-\half\varphi_{\rm gh}(z)}&(4.30a)\cr
V^{(2)}_{-\half}(k,z) &= u^{2\dot\alpha} (k) S_{\dot\alpha} (z) 
e^{ik_\mu X^\mu (z)} S_{A} (z) e^{{-Q}\phi(z)} 
S_{\dot n} (z) 
e^{-\half\varphi_{\rm gh}(z)}&(4.30b)\cr}$$
where $k_\mu k^\mu = 0$;
$S_\alpha$, $S_{\dot\alpha}$ are the spin fields of 4-dimensional spacetime,
and $S_n$, $S_{\dot n}$, $n=1,\dot n = \dot 1$ are the 
weight ${\textstyle{1\over 8}}$ spin fields
associated with the fermionic component $\psi^\ell$ of the supercoordinates
on $T^{(2)}$. 
In (4.30), the two linearly independent solutions to the massless
Dirac equation $k\dt\gamma u(k) = 0$ are given by solutions of the
Weyl equations
$$k_\mu\sigma^{\mu\dot\alpha}_{\hskip8pt\beta} u^{1\beta} = 0\;\hskip35pt
k_\mu\bar\sigma^{\mu\alpha}_{\hskip8pt\dot\beta} u^{2\dot\beta} = 0\,
\eqno(4.31)$$
and $\bar\sigma^0 = -\sigma^0$, $\bar\sigma^i = \sigma^i$.

{}From (4.25,4.30), we see that the Ramond states
$V^{(1,2)}_{-\half} (k,0) |0\rangle$ each correspond to a doublet
under the level one $SU(2)$ algebra $\bar {\cal S}_0^i$. 
Since
$$\eqalignno{iT^i(z) S_A(\zeta) &= (z-\zeta)^{-1} Q {i\over 4}\epsilon_{ijk}
(\gamma^j\gamma^k)_A^{\hskip 3pt B} S_B(\zeta)\cr
iT^i(z) S_{\dot A}(\zeta) &= (z-\zeta)^{-1} Q {i\over 4}\epsilon_{ijk}
(\gamma^j\gamma^k)_{\dot A}^{\hskip 3pt\dot B} S_{\dot B}(\zeta)
&(4.32)\cr}$$
we have 
$$\eqalignno{i T_0^3 S_1(0) |-{\scriptstyle Q}\rangle 
= - Q\hhalf S_1(0) |-{\scriptstyle Q}\rangle,&\qquad
i T_0^3 S_2(0) |-{\scriptstyle Q}\rangle 
= Q\hhalf S_2(0) |-{\scriptstyle Q}\rangle\,;\cr
i T_0^3 S_{\dot 1} (0) |0\rangle = -Q\hhalf S_{\dot 1} (0) |0\rangle,&\qquad
i T_0^3 S_{\dot 2} (0) |0\rangle = Q\hhalf S_{\dot 2} (0) |0\rangle\,,
&(4.33)\cr}$$
so that
$V^{(1,2)}_{-\half} (k,0) |0\rangle$ each also correspond to a doublet
under the $SU(2)$ algebra $T_0^i$.

Therefore these fermion states, which are highest weight states of 
massless representations of the $N=4$ algebra and consequently
of the $N=1$ subalgebra, carry charge under an 
$SU(2)$ gauge algebra. In the next section, we 
discuss the implications of this for
representations of the $N=1$ super Kac-Moody algebra.

\vskip15pt
\centerline{\bf 5. Fivebrane representation of $N=1$ SVA and SKMA}
\vskip5pt
The construction of the fivebrane $c=6$, $N=1$
superVirasoro algebra (SVA) (which is a subalgebra of the $N=4$ system)
given by 
$$\eqalignno
{L(z)&= -{1\over 2}
J^0 J^0 - {1\over {k+2}} J^i J^i -{1\over 2}\partial\psi^A \psi^A
+\delta {1\over 4 z^2} + {1\over 2} Q\partial J^0&(5.1)\cr
F(z)&= \psi^0 J^0 -Q \psi^i J^i
-Q{\textstyle{1\over 6}}
\epsilon_{ijk}\psi^i\psi^j\psi^k - Q\partial\psi^0&(5.2)\cr}$$
for $\delta = 0,1$ for NS or R sector respectively, together with
the weight one-half superfield whose components are
$$\psi^i(z) \,,\qquad 
T^i (z) = Q [ J^i + \hhalf\epsilon_{ijk}\psi^j\psi^k ]\,\eqno(5.3)$$ 
form a representation of the super $SU(2)$ Kac-Moody algebra (SKMA) 
mixing with the $N=1$ SVA:
$$\eqalignno{F(z) \psi^i(\zeta)&= (z-\zeta)^{-1}
T^i(\zeta)\,,\quad
F(z) T^i(\zeta) = (z-\zeta)^{-2} \psi^i(\zeta)
+ (z-\zeta)^{-1}\partial\psi^i(\zeta)\,&(5.4)\cr
T^i(z) T^j(\zeta) &= -{\delta_{ij}\over (z-\zeta)^2} + 
{Q\epsilon_{ijk}T^k(\zeta)\over (z-\zeta)}\cr
T^i(z) \psi^j(\zeta) &= {Q\epsilon_{ijk} \psi^k(\zeta)\over (z-\zeta)}\,,
\qquad
\psi^i(z) \psi^j(\zeta) = -{\delta_{ij}\over (z-\zeta)}\,&(5.5)\cr}$$
where $J^i$ is the  level $k$ SU(2) KMA and $Q = -{\sqrt{2\over (k+2)}}$. 
We note that superfield associated with $\psi^0(z)$ has lower and
upper components
$\psi^0(z) e^{-Q\phi(z)}\,,\qquad
T^0 (z) = -{1\over Q} \partial_z e^{-Q\phi(z)}$ so that unless $Q=0$
the current $T^0(z)$ is not the generator of a $U(1)$ Heisenberg algebra. 

For simplicity, we can consider $k=0$, in which case $Q=-1$ and the
currents $J^i$ decouple from the system, so that
(5.1-5.5) become with $c=6$: 
$$\eqalignno
{&L(z) = -{1\over 2}
J^0 J^0 -{1\over 2}\partial\psi^A \psi^A
+\delta {1\over 4 z^2} - {1\over 2} \partial J^0\cr
&F(z) = \psi^0 J^0 
+ {\textstyle{1\over 6}}
\epsilon_{ijk}\psi^i\psi^j\psi^k + \partial\psi^0\cr
&\psi^i (z) \,,\qquad                 
T^i (z) = - \hhalf\epsilon_{ijk}\psi^j\psi^k \,\cr
&F(z) \psi^i(\zeta)= (z-\zeta)^{-1}
T^i(\zeta)\,,\quad 
F(z) T^i(\zeta) = (z-\zeta)^{-2} \psi^i(\zeta) 
+ (z-\zeta)^{-1}\partial\psi^i(\zeta)\,\cr 
&T^i(z) T^j(\zeta) = -{\delta_{ij}\over (z-\zeta)^2} - 
{\epsilon_{ijk}T^k(\zeta)\over (z-\zeta)}\cr       
&T^i(z) \psi^j(\zeta) = -{\epsilon_{ijk} \psi^k(\zeta)\over (z-\zeta)}\,,
\qquad
\psi^i(z) \psi^j(\zeta) = -{\delta_{ij}\over (z-\zeta)}\,.&(5.6)\cr}$$
For the system (5.6), we find that the Ramond states
$S_A(0) e^{-Q\phi(0)} |0\rangle$ and $S_{\dot A}(0) |0\rangle$ satisfy
$$\eqalignno{F_0\, S_A(0) |-{\scriptstyle Q}\rangle &= 0\,,\quad
F_0 S_{\dot A} (0)| 0\rangle = 0&(5.7)\cr}$$
but that they are carry non-trivial $SU(2)$ charge:
$$\eqalignno{&-i\{F_0, \psi_0^3\}\,  S_1(0) |-{\scriptstyle Q}\rangle = 
-i T_0^3\, S_1(0) |-{\scriptstyle Q}\rangle = -\hhalf S_1(0) 
|-{\scriptstyle Q}\rangle&(5.8a)\cr 
&-i \{F_0, \psi_0^3\}\,  S_2(0)  
|-{\scriptstyle Q}\rangle =
-i T_0^3 \,S_2(0)  
|-{\scriptstyle Q}\rangle 
= \hhalf S_2(0) |-{\scriptstyle Q}\rangle&(5.8b)\cr 
&-i \{F_0, \psi_0^3\}\,  S_{\dot 1}(0) |0\rangle =
-i T_0^3 S_{\dot 1}(0) |0\rangle = -\hhalf S_{\dot 1}(0) |0\rangle&(5.8c)\cr 
&-i \{F_0, \psi_0^3\}  S_{\dot 2}(0) |0\rangle =
-i T_0^3\, S_{\dot 2} (0) |0\rangle = +\hhalf S_{\dot 2}(0) 
|0\rangle\,.&(5.8d)\cr} $$ 
Since these Ramond states are massless representations
of the $N=4$ algebra, they are also representations of the 
$N=1$ subalgebra given in (5.6). 
In this way, in a closed string theory, a non-abelian $SU(2)$ gauge symmetry
can occur on either side which supports massless charged fermions. 
This evades the argument of the DKV theorem\ref{18},
which had prevented massless (R,NS) fermions in a string model with a 
left-moving super Kac-Moody algebra. Furthermore these fermions carry
left-moving $SU(2)$ charge.
Given (5.16) and (2.19), these states have positive norm.

The mechanism which allows for massless fermions charged under an
$SU(2)$ gauge symmetry occurring in the Ramond sector is the choice of 
inner product (2.19) and it works as follows: 
{}From (5.7),(5.8a) we have 
$$ i F_0 \psi_0^3\,  S_1(0) |-{\scriptstyle Q}\rangle =
i T_0^3\, S_1(0) |-{\scriptstyle Q}\rangle 
= \hhalf S_1(0) |-{\scriptstyle Q}\rangle\eqno(5.9)$$
since $F_0 S_1(0) |-{\scriptstyle Q}\rangle = 0$ from (5.7). 
As in (5.16), taking the norm of the righthand side of (5.9), we have
$$ \hhalf \lim_{z\rightarrow \infty}
z^{\half} \langle 0 | S^1(z) S_1(0) |-{\scriptstyle Q}\rangle 
= \hhalf\,.\eqno(5.10)$$
It follows that this action on the lefthand side of (5.9) is also non-zero:
$$ \eqalignno{\lim_{z\rightarrow \infty}
z^{\half} \langle 0 |S^1(z) F_0 i\psi_0^3\,  
S_1(0) |-{\scriptstyle Q}\rangle &=
\lim_{z\rightarrow \infty}
z^{\half} \langle 0 |S^1(z) F_0 {\textstyle(-{1\over\sqrt 2})}
S_{\dot 1}(0) |-{\scriptstyle Q}\rangle = \hhalf&(5.11)\cr}$$
so
$$\lim_{z\rightarrow\infty} z^\half \langle 0 |S^1(z) F_0
= {\textstyle(-{1\over\sqrt 2})} 
\lim_{z\rightarrow\infty} z^\half \langle 0 |S^{\dot 1}(z)\ne 0\eqno(5.12)$$
which is the same statement as 
$$ [\,\lim_{z\rightarrow\infty} z^\half\langle 0 | S^1(z) F_0 \,\, ]^\dagger 
= {\textstyle(-{1\over\sqrt 2})} S_{\dot 1}(0) |-{\scriptstyle Q} \rangle
\ne 0\,.\eqno(5.13)$$
Therefore, although
$$F_0 S_1(0) |-{\scriptstyle Q}\rangle = 0\,,\eqno(5.14)$$
we have that $F_0$ {\it acting on the state adjoint to} 
$S_1(0) |-{\scriptstyle Q}\rangle$ {\it is not zero}, 
as shown in (5.12),(5.13). This is allowed, since although
the hermiticity property of the inner product requires
$$ \eqalignno
{0&= (\,S_1(0) |-{\scriptstyle Q}\rangle , F_0 S_1(0) 
|-{\scriptstyle Q}\rangle\, )
= (\,[\,\lim_{z\rightarrow\infty} 
z^\half \langle 0 | S^1(z) F_0 \,\, ]^\dagger, 
S_1(0) |-{\scriptstyle Q}\rangle\, )&(5.15a)\cr
=&\lim_{z\rightarrow \infty} z^{\half}
\langle 0 | S^1(z) F_0 S_1(0)|-{\scriptstyle Q}\rangle 
=-{\scriptstyle{1\over{\sqrt 2}}}
\lim_{z\rightarrow \infty} z^{\half}\langle 0| S^{\dot 1} (z)
S_1(0)|-{\scriptstyle Q}\rangle = 
-{{\delta^{\dot 1}_1}\over{\scriptstyle\sqrt 2}} 
=0\,,\qquad &(5.15b)\cr}$$
the righthand side of (5.15a) {\it vanishes because
the states in the inner product are orthogonal, not because}
$[\,\lim_{z\rightarrow\infty} z^\half \langle 0 | S^1(z) F_0 \,\,]^\dagger$ 
{\it is zero}.

The norm of the spin field states $|\psi\rangle\equiv
S_{A}(0)|0\rangle $ without the Liouville
contribution is given by\ref{19} 
$$\eqalignno{\langle\psi | \psi\rangle &= \lim_{{z\rightarrow \infty}\atop
{\zeta\rightarrow 0}}
z^{2h} \langle 0 | V(\bar\psi, z) V(\psi,\zeta)|0\rangle
= \lim_{{z\rightarrow \infty}\atop
{\zeta\rightarrow 0}}
z^{\half} \langle 0 | S^A (z) S_A (\zeta)|0\rangle\cr               
&= \lim_{{z\rightarrow \infty}\atop
{\zeta\rightarrow 0}}
z^{\half} (z-\zeta)^{-\half} C^{AB} C^{-1}_{BA}
= \delta^A_{\hskip3pt A} = 1\,,&(5.16)\cr}$$
where the vertex for the state $\bar\psi$ conjugate to $\psi$ is given by
$V(\bar\psi, z)\equiv z^{-2h} V(\psi, {1\over{z^\ast}})^\dagger$.
Since $S_{An}^\dagger = S^A_{-n}$, the state conjugate to 
$S_A(0)|0\rangle$ is $S^A(0)|0\rangle$. 
We have\ref{19} 
$$\eqalignno{ \psi = \lim_{z\rightarrow 0} V(\psi, z) |0\rangle\,,\qquad
&\bar\psi = \lim_{z\rightarrow 0} V(\bar\psi, z) |0\rangle\cr
\langle \bar\psi | =\lim_{z\rightarrow\infty} z^{2h} \langle 0 | 
V(\psi, z)\,,\qquad
&\langle \psi | =\lim_{z\rightarrow\infty} z^{2h} \langle 0 |
V(\bar\psi, z)\,.&(5.17)\cr}$$
For the discrete Liouville states $|q\rangle = e^{q\phi(0)}|0\rangle$,
the non-vanishing inner product is\break
$\langle -q-Q| q\rangle = 1$. 
As in sect. 2, we define the adjoint of $|q\rangle = e^{q\phi(0)}|0\rangle$
to be $\langle -q-Q|$, so that the state
$|q\rangle$ has non-zero norm: $|| |q\rangle ||^2
= \langle -q-Q | q\rangle = 1$.
Clearly all propagating states, i.e. those
which have continuous momenta such as (4.30) with 
$k_\mu k^\mu = 0$ but $k^\mu\ne 0$, must have positive norm.
This completes our discussion of the existence of charged massless Ramond
states. ( In (5.6)-(5.15) we choose $Q=-1$. The discussion is 
straightforwardly generalized for any real $Q = -{\sqrt{2\over {k+2}}}$. 
The limit $Q\rightarrow 0$ collapses to zero charge fermions.) 

The set of consistent GSO projections for states such as (4.30) in  
a string theory with space-time perturbative unitarity 
is being investigated. This may include projecting away massive
states such as (4.16) which depend on a continous parameter $y$.\ref{20} 
We note here that 
another set of BRST invariant Ramond operators is given by
$$\eqalignno{ {V'}^{(1)}_{-\half}(k,z)
&= [u^{1\alpha} (k) S_\alpha (z) 
S_A(z) e^{{-Q}\phi(z)} - Q v^{1\dot\alpha} (k) S_{\dot\alpha} (z) 
\gamma_A^{\hskip3pt 0\dot A} S_{\dot A}(z) ]\cr
&\hskip 10pt \cdot S_n(z)\,e^{ik_\mu X^\mu (z)}  
e^{-\half\varphi_{\rm gh}(z)}&(5.18)\cr}$$ with a similar expression for
${V'}^{(2)}_{-\half}(k,z)$. 
These vertex operators, associated with different projections from (4.30),
give rise to non-abelian tree amplitudes of Ramond-Ramond states.\ref{21}  
Here we defined two additional spinors 
$v^\ell(k)$ by $k\dt\gamma v^\ell \sim u^\ell$, i.e.
$${\textstyle{1\over\sqrt 2}}
k_\mu\bar\sigma^{\mu\alpha}_{\hskip8pt\dot\beta} v^{1\dot\beta}
= u^{1\alpha}\,\;\hskip35pt
{\textstyle{1\over\sqrt 2}}
k_\mu\sigma^{\mu\dot\alpha}_{\hskip8pt\beta} v^{2\beta}
= - u^{2\dot\alpha}\,.\eqno(5.19)$$  

\vskip15pt
\centerline{\bf 6. Conclusions}
\vskip5pt
\parskip=6pt 
The specific fivebrane CFT permits us to compute the spectrum of 
string excitations around the corresponding background solution. In this
sense, the choice of CFT provides non-perturbative information.

In formulations of string theory, the question arises as to whether 
1) all the necessary information  
can be seen as the spectrum of some CFT, or instead 
2) does a framework broader than CFT need to 
be employed to realize the natural string ground state.
In this paper, 
we investigate only the first point of view, but we point to 
effects which occur in CFTs, albeit unconventional ones with background
charge, which lead to novel gauge properties of the spectrum.  
This is in the spirit that in order to solve any non-linear system, 
ultimately
a perturbation theory about some new point will do the job. 
In this paper, that new point is the Type II superstring 
with an internal background charge CFT.

The fact that in physics the massless spectrum, i.e. the low mass spectrum
carries representations of the gauge group, and that the gauge generators
are the zero modes of weight one conformal fields, suggests that
at least the massless states should be described by some CFT. 

In this paper, we were led to consider a conformal field theory with
background charge in order to have charged Ramond states. The fivebrane
conformal field theory provides a specific example of such a conformal
field theory, and its use was motivated
by the fact that the fivebrane, an NS-NS brane, carries charge under the
NS gauge bosons. Furthermore, a connection between gauged supergravity 
theories\ref{22-26} which necessarily 
have tree level cosmological constants,
and string theories\ref{21,11} serves to increase 
interest in pursuing non-critical string theories\ref{10,11}.
The latter provide a link between string theory and 
field theories with tree level cosmological constants.

The presence of gauge symmetry in string theory is closely tied to 
affine Kac-Moody algebras\ref{27-29}.
We showed in sect.5 how a choice of inner product for the discrete Liouville
states in this CFT leads to a generalization of representations of the 
$N=1$ super $SU(2)$ Kac-Moody algebra. In principle,
this now allows an $SU(2)$ non-abelian factor in the gauge group on one
side, with the original non-abelian factors (like $SU(3)\otimes U(1)$)
coming from the other side, and provides symmetry enhancement within
the economical Type II superstring.  

The author acknowledges valuable discussions with 
A. Belopolsky, R. Blumenhagen, M. Flohr, P. di Francesco, and P. Goddard, 
and is grateful for the hospitality of St.John's College, Cambridge
where part of this work was done. 

\vskip15pt
\vfill\eject
\centerline{\bf Appendix}
\vskip10pt
\leftline{$N=4$, $c=6$ SUPERCONFORMAL ALGEBRA}

The currents of (2.5) close on the $N=4$ superconformal algebra
with $c=6$ which is given by 
$$\eqalignno{&L(z) L(\zeta) = {{c\over 2}\over (z-\zeta)^4}
+ {2L(\zeta)\over{(z-\zeta)^2}} + {\partial L(\zeta)\over (z-\zeta)}\cr
&L(z) F(\zeta) = {{3\over 2}F(\zeta)\over (z-\zeta)^2} +
{\partial F(\zeta)\over(z-\zeta)}\cr
&L(z) F^i(\zeta) = {{3\over 2}F^i(\zeta)\over (z-\zeta)^2} + 
{\partial F^i(\zeta)\over (z-\zeta)}\cr
&L(z) S^i(\zeta) = {S^i(\zeta)\over (z-\zeta)^2} +
{\partial S^i(\zeta)\over (z-\zeta)}\cr
&F(z) F(\zeta) = {{2c\over 3}\over (z-\zeta)^3} +
{2L(\zeta)\over{(z-\zeta)}}\cr
&F(z) F^i(\zeta) = - {4 S^i(\zeta)\over (z-\zeta)^2} 
-  {2\partial S^i(\zeta)\over{(z-\zeta)}}\cr
&F^i(z) F^j(\zeta) = {\delta^{ij}{2c\over 3}\over (z-\zeta)^3}
- {4 \epsilon_{ij\ell} S^\ell(\zeta)\over (z-\zeta)^2} 
- {2 \epsilon_{ij\ell} \partial S^\ell (\zeta)\over{(z-\zeta)}}  
+ {2 \delta^{ij} L(\zeta)\over{(z-\zeta)}}\cr
&S^i(z) S^j(\zeta) = -{n\delta^{ij}\over 2 (z-\zeta)^2}            
+ {\epsilon_{ij\ell} S^\ell(\zeta)\over (z-\zeta)}\cr 
&S^i(z) F(\zeta) =  { F^i(\zeta)\over 2 (z-\zeta)}\cr
&S^i(z) F^j(\zeta) = {1\over (z-\zeta)} 
[ -\delta^{ij} F(\zeta) + \epsilon_{ij\ell} F^\ell (\zeta) ]\,.&(A.1)\cr}$$
The central charge $c$ and the level $n$ of the $SU(2)_n$ currents
$S^i$ are related by $c = 6n$. The condition $c=6$ sets
$S^i$ at level one.  

We define complex supercurrents as
$$\eqalignno{{\cal G}^1 &\equiv {F - i F^3\over \sqrt 2}\,,\qquad
{\cal G}^2 \equiv {F^2 - i F^1\over \sqrt 2 }\cr
\bar{\cal G}^1 &\equiv {F + i F^3\over \sqrt 2}\,,\qquad
\bar{\cal G}^2 \equiv {F^2 + i F^1\over \sqrt 2}\cr}$$
and note that ${\cal G}^1,{\cal G}^2$ transform as a doublet under
$S^i$ etc.
\vfill\eject
\centerline{\bf References}
\vskip 5pt
\item{ 1.}  A. Strominger, 
{\it Heterotic solitons}, Nucl. Phys. {\bf B343} (1990) 167.
\item{ 2.} C. Callan, J. Harvey, and A. Strominger, 
{\it Worldsheet approach to heterotic
instantons and solitons}, Nucl. Phys. {\bf B359} (1991) 611; and
{\it Worldbrane actions for string solitons}, Nucl. Phys. 
{\bf B 367} (1991) 60.
\item{ 3.} M.Duff, R. Khuri, and J.X. Lu, {\it String solitions}, 
Phys. Rep. {\bf 259} (1995) 213-326, hep-th/9412184.
\item{ 4.} C. Callan, {\it Instantons and solitons 
in heterotic string theory},
Swieca Summer School, June 1991; hep-th/9109052.                             
C. Callan, J. Harvey, and A. Strominger, {\it Supersymmetric string solitions},
1991 Trieste Spring School on String Theory and Quantum Gravity,             
hep-th/9111030.
\item{ 5.} A. Sevrin, W. Troost, and A. van Proyen,  
{\it Superconformal algebras
in two dimensions with $N=4$}, Phys. Lett. {\bf B208} (1988) 447.
\item{ 6.}  C. Hull and P. Townsend, Nucl. Phys. {\bf B438} (1995) 109,
hep-th/9410167; Nucl. Phys. {\bf B451} (1995) 525,
hep-th/9505073.
\item{ 7.} E. Witten, {\it String theory in various dimensions},
Nucl. Phys. {\bf B443} (1995) 85, hep-th/9503124.
\item{ 8.} J. Polchinski, {\it Dirichlet-Branes and Ramond-Ramond Charges},
Phys. Rev. Lett. {\bf 75} (1995) 4727, hep-th/9510017.
\item{ 9.} P. Townsend, {\it D-branes from M-branes}, Phys. Lett. 
{\bf B373} (1996) 68, hep-th/9512062.
\item{10.} N. Seiberg and D. Kutasov, {\it Non-critical superstrings},
Phys. Lett. {\bf B251} (1990) 67.
\item{11.} I. Antoniadis, S. Ferrara, C. Kounnas, 
{\it Exact supersymmetric string
solutions in curved gravitational backgrounds},
Nucl. Phys.{\bf B421} (1994) 343, hep-th/9402073.
\item{12.} T. Eguchi and A. Taormina, {\it Unitary representations of the
$N=4$ superconformal algebra}, Phys. Lett. {\bf B196} (1986) 75;
{\it Character formulas for the $N=4$ superconformal algebra},
Phys. Lett. {\bf B200} (1988) 315.
\item{13.} G. Mostow and J. Sampson, {\it Linear Algebra},
New York: McGraw-Hill, 1969.
\item{14.} T. Banks, J. Dixon, D. Friedan, and E. Martinec,
Nucl. Phys. {\bf B299} (1988) 613.
\item{15.} T. Banks, J. Dixon, Nucl. Phys. {\bf B307} (1988) 93.
\item{16.} D. Friedan, E. Martinec, and S. Shenker, Nucl. Phys. {\bf B271}
(1986) 93.
\item{17.} I. Klebanov and A. Polyakov, Mod. Phys. Lett A6(1991)635
[hep-th/9109032];
I. Klebanov and A Pasquinucci, hep-th/9210105.
\item{18.} L. Dixon, V. Kaplunovsky and C. Vafa, Nucl. Phys.
{\bf B294} (1987) 43.
\item{19.} R. Bluhm, L. Dolan and P. Goddard, {\it Conformal field theory
of twisted vertex operators}, Nucl. Phys. {\bf B338} (1990) 529.
\item{20.} L. Dolan, work in progress.
\item{21.} L. Dolan and S. Horvath, Nucl. Phys. {\bf B448} (1995) 220.
\item{22.} E. Cremmer and B. Julia, ``{\it The SO(8) Supergravity}'',
Nucl. Phys.{\bf B159} (1979) 141.
\item{23.} B. de Wit and H. Nicolai, ``{\it $N=8$ supergravity with local
$SO(8)\otimes SU(8)$ invariance}'', Phys. Lett. {\bf B108} (1982) 285;
``{\it $N=8$ Supergravity}'', Nucl. Phys.{\bf B208} (1982) 323.
\item{24.} M.Duff, B. Nilsson, and C. Pope, 
``{\it Kaluza-Klein Supergravity}'', Phys. Rep. {\bf 130} (1986) 1.
\item{25.} L. Romans, Phys. Lett. {\bf B169} (1986) 374.
\item{26.} I. Antoniadis, C. Bachas, and A. Sagnotti, Phys. Lett. {\bf B235}
(1990) 255.
\item{27.} R. Bluhm, L. Dolan and P. Goddard, Nucl. Phys.
{\bf B289} 364 (1987); Nucl. Phys. {\bf B309} 330 (1988).
\item{28.} M.B. Green, J. Schwarz and E. Witten, {\it Superstring
theory} (Cambridge University Press, 1987).
\item{29.} L. Dolan, ``The Beacon of Kac-Moody Symmetry for Physics'',
Notices of the American Mathematical Society, {\bf 42} No. 12, 1489-1495,
December 1995. 
\vfill\eject
 
\bye